\documentclass[a4paper, usenatbib]{mnras}

\usepackage{graphicx}

\usepackage{epsfig}

\usepackage{subfig}

\usepackage{multirow}

\usepackage{mathrsfs,amsmath} 

\usepackage[utf8]{inputenc}

\usepackage{epsfig}
\usepackage{epstopdf}
\usepackage{subfloat}
\usepackage{hyperref}
\usepackage{multirow}
\usepackage{graphicx,times}  
\usepackage{soul}
\usepackage{amssymb,amsmath}

\def\araa{ARAA}
\def\mnras{MNRAS}
\def\apj{APJ}
\def\aj{AJ}
\def\apjl{APJL}
\def\aap{AAP}

\def\na{NewAstro}
\def\ell{Elsevier}

\newcommand{\id}{{\rm d}}

\newcommand{\Gt}{G(\vt)}
\newcommand{\dn}{\delta n(\vt)}
\newcommand{\lsf}{\phi(\vt, \nu)}
\newcommand{\vL}{v_L(\vt)}
\newcommand{\vR}{v^{\Omega}(\vt)}
\newcommand{\vT}{\delta v^{T} (\vt)}

\newcommand{\vt}{{\vec \theta}}
\newcommand{\vU}{{\vec{ U}}}

\def\HI{{H~{\sc i} }}

\begin{document}

\title[Large scale turbulence cascade in NGC~5236]{Evidence of large scale energy cascade in the spiral galaxy NGC~5236}
\author[Meera Nandakumar and Prasun Dutta ]
{Meera Nandakumar$^{1}$\thanks{Email: meeranandkr.rs.phy17@itbhu.ac.in}, 
Prasun Dutta$^{1}$\thanks{Email:pdutta.phy@itbhu.ac.in},  
\\$^{1}$ Department of Physics, IIT (BHU) Varanasi, 221005  India. 
}
\maketitle 

\begin{abstract}
	Turbulence plays an important role in the structure and dynamics of the galaxy and influences various processes therein including star formation. In this work, we investigate the large scale turbulence properties of the external spiral galaxy NGC~5236. We combine the  VLA multi-configuration archival data with the new GMRT observation to estimate the column density and line of sight velocity fluctuation power spectra for this galaxy over almost two decades of length scales. The energy input scale to the ISM turbulence is found to be around $6$ kpc. Power spectra of the two-dimensional turbulence in the galaxy's disk follow a  power law with a slope $-1.23\pm0.06$ for the column density and $-1.91\pm0.08$ for the line of sight velocity. The measured power spectra slopes strongly suggest in favour of a compressive forcing with a steady energy input of $\sim 7 \times 10^{-11}$ erg cm$^{-2}$ sec$^{-1}$. We conclude that much of these originate from the gravitational instabilities and self-gravity in the disk. This is the first and most comprehensive study of turbulence statistics for any external spiral galaxy.
\end{abstract}


\section{Introduction}
The observed scale-invariant structures in the interstellar medium (ISM) of the galaxies are often attributed to compressible fluid turbulence. The energy input to the ISM turbulence comes from various sources, shocks in the spiral arms, rotational shear, non-circular motion from the star formation feedback, from the supernovae and  proto-stellar etc \citep{2004ARAA..42..211E,2020arXiv200301131U}. Density and velocity fluctuations in the ISM turbulence can be quantified using estimators of the  two-point statistics of  the spatial intensity distribution of any line emission \citep{2000ApJ...537..720L,2001ApJ...555L..33P,2010ApJ...708.1204B,2004ApJ...616..943L,2002ApJ...566..289B}. A commonly used  quantifier of the turbulence statistics is the power spectrum of the special intensity fluctuation, where the column density power spectrum and in some cases the line of sight velocity power spectrum can be inferred. For compressible fluid turbulence, these power spectra are expected to follow power laws \citep{2009ApJ...692..364F, 1996ApJ...458..739F}  with the amplitude and the slope having information about the nature of the turbulence and its source.

Much of the information about ISM turbulence in the Galaxy is known from the measurements of \HI power spectra  in emission \citep{1983AA...122..282C,1993MNRAS.262..327G} and absorption \citep{2000ApJ...543..227D, 2010MNRAS.404L..45R}. \citet{2001ApJ...548..749E,2015ApJ...810...33C,2006MNRAS.372L..33B} have estimated the power spectra of \HI intensity fluctuations for the Large and Small Magellanic Clouds,  DDO~210  and inferred existence of  large scale coherent structures in those dwarf galaxies suggesting long-range turbulence energy cascade. \citet{2009MNRAS.398..887D} and later \citet{2012ApJ...754...29Z, 2017AJ....153..163M} have found that the large scale turbulence cascade exists in a large number of dwarfs.  This energy cascade by turbulence is expected to affect the fragmentation and the collapse of the \HI and later molecular clouds in the star formation process and moderate the rate of star formation in  galaxies \citep{2004RvMP...76..125M}.  \citet{2008MNRAS.384L..34D} estimated the \HI power spectra of the external spiral galaxy NGC~628, where they find that the power spectra follow a power law with a slope of -1.6 over a length scale ranging from $800$ pc to $8$ kpc indicating a large scale energy input in the turbulence cascade.  This was followed by measurement of \HI intensity fluctuation power spectra of the nearly face-on galaxy NGC~1058 \citep{10.1111/j.1745-3933.2009.00684.x} and the harassed galaxy  NGC~4254 \citep{2010MNRAS.405L.102D} in the virgo cluster. In more comprehensive studies \citet{2013MNRAS.436L..49D,2013NewA...19...89D} 
  observed that collectively for $18$ spiral galaxies in their sample the intensity fluctuation power spectra follow power law over a range of $300$ pc to $16$ kpc. They find that half of the galaxies in their sample follow a power law with a slope in between $-1.5$ and $-1.9$. 

Most of the studies listed above estimate the \HI specific intensity fluctuation power spectra,  which is directly related to the \HI column density power spectra \citep{2013MNRAS.436L..49D}. The range of length scales over which the density fluctuation power assumes a power law provides us with the range of turbulence cascade, the injection scale of energy and to some extent the nature of turbulence forcing \citep{2009ApJ...692..364F}. However, to infer on the momentum and energy input and the rate of energy transfer by the turbulence cascade the velocity statistics of the turbulence is essential. Spectral line observations, like the \HI  let us estimate the line of sight velocity statistics. In this regard line of sight velocity dispersion  are investigated in literature \citep{2009AJ....137.4424T,2015AJ....150...47I}. However, the  two-point statistics of the velocity fluctuations carry more information about the turbulent dynamics. Several estimators of the two-point statistics of the turbulent velocity fluctuations are devised \citep{Esquivel_2005,2006ApJ...652.1348L,2000ApJ...537..720L,2001ApJ...555L..33P} in literature and are successfully used for estimating the velocity power spectra of the Galaxy and the nearby dwarf galaxies \citep{2017ApJ...845...53N,
2012ApJ...754...29Z}. These studies find evidence of large scale driving for the ISM turbulence in Small Magellanic Cloud (SMC) and Large Magellanic Cloud (LMC) \citep{2019ApJ...887..111S, 2015ApJ...810...33C, 2001ApJ...551L..53S} with injection scale as high as $2$ kpc generating supersonic turbulence. \citet{2015MNRAS.452..803D} discussed the limitations of most of these techniques to estimate the turbulence velocity power spectra for the external spiral galaxies. An alternative estimator for the column density as well as the line of sight velocity power spectrum using radio interferometric observations of \HI fluctuations is proposed by \citet{2016MNRAS.456L.117D}. In this work, we present an implementation of this visibility moment estimator (VME hence forth) and use it to derive ISM turbulence properties of the external spiral galaxy  NGC~5236. 

NGC~5236 (Messier 83) is a SAB(s)c type almost face on spiral galaxy \citep{1991rc3..book.....D}. \citet{1973A&A....29..425B} have used the  Nancay radio telescope to obtain a brightness temperature map for this galaxy. A large fraction of the \HI in this galaxy is found by \citet{1981A&A...100...72H} to reside outside the Holmberg radius. \citet{2004A&A...422..865L} have found using CO observation that at least a part of the NGC~5236's disk is gravitationally unstable. \citet{2008AJ....136.2563W} have observed this galaxy using multiple Very Large Array (VLA) \footnote{NRAO Very Large Array}  configuration as a part of The HI Nearby Galaxy Survey (THINGS). \citet{2013NewA...19...89D} have estimated the \HI specific intensity fluctuation power spectra of this galaxy and found it to follow a power law of slope $-1.9 \pm 0.2$ over a length scale ranging from $800$ pc to $7.5$ kpc using the data from THINGS survey. In this work, we combine the VLA multi-configuration data from the THINGS survey with the new uGMRT\footnote{uGMRT: upgraded Giant metrewave radio telescope, Pune} \citep{1991ASPC...19..376S,2017CSci..113..707G} observation and  use the VME to measure the column density and line of sight velocity fluctuation in NGC~5236. We infer about the type of turbulence forcing and energy input at the driving scale. Rest of the paper is arranged in the following way. In section~2 we discuss the VME, the observation and analysis details are discussed in section~3, section~4 details the results and we conclude in section~5.

\section{Visibility Moment Power Spectrum estimation}
\subsection{Specific intensity of \HI emission}
The visibility moment estimator (VME) for column density and turbulence velocity power spectrum is introduced in \citet{2016MNRAS.456L.117D}. We use a variant of the VME for this work.  The angular extent of the galaxy is rather small and we consider a flat sky approximation here.  We use a cartesian coordinate system with its origin at  the centre of the galaxy of interest. The tangent plane about that origin in the sky has the coordinates $\theta_x$ and $\theta_y$. The line of sight direction gives the  third axis of the cartesian system and we call it by $z$. Any position $P$  in the sky is marked by the vector $\vt = [\theta_x, \theta_y]$ in the tangent plane. Here $[\theta_x, \theta_y]$ gives the angular separation of the point P in two orthogonal directions in the tangent plane with respect to the origin.   The specific intensity of \HI 21-cm radiation from an external galaxy can be written as \citep{2011piim.book.....D}
\begin{equation}
I(\vt, \nu) = \frac{3 h \nu_0 A_{21}}{16 \pi} \int d{\rm z}\  n_{\rm HI} (\vt, z) \phi(\vt, z, \nu).
\end{equation}
The constants $\nu_0, A_{21}$ are the \HI line frequency and corresponding Einstein's coefficient.  The number density of neutral hydrogen $n_{HI}$ can be written as 
\begin{equation}
n_{\rm HI} (\vt, z) = n_0\ G(\vt, z)\, \left [1 + \delta n (\vt, z) \right ],
\end{equation}
where $n_0$ is the average \HI number density over the entire galaxy. Here we consider the centre of the galaxy and the tangent plane coincides. In general, the galactic disk can have an arbitrary orientation with respect to the tangent plane we have considered here. The galaxy \HI profile function $G(\vt, z)$ times $n_0$ gives the large scale variation of the \HI  number density in the coordinate system discussed here, while $\int \id \vt \int \id z G(\vt, z) = 1$.  Note that for any galaxy not oriented exactly face on, a circular disk will be seen in projection in the tangent plane.   The term $\delta n$ gives the random variation in the \HI number density. For  a nearly face-on thin disk (disk thickness $l$), the specific intensity of  \HI emission can be written as
\begin{equation}
I(\vt, \nu) = I_0 \ G(\vt)\, \left [1 + \delta n (\vt) \right ] \phi(\vt, \nu).
\end{equation}
Here $I_0 = \frac{3 h \nu_0 A_{21} n_0 l }{16 \pi}$ gives the total \HI flux from the galaxy, the quantities $\Gt = G(\vt, z=0), \dn = \delta n(\vt, z=0)$ and $\lsf = \phi(\vt, z=0, \nu)$.
The line shape function $\lsf$ can be modelled as
\begin{equation}
\lsf = \frac{1}{\sqrt{2 \pi} \sigma_{\nu}} \exp \left [ - \frac{1}{2} \left ( \frac{\nu - \nu_0 ( 1 - \frac{v_L(\vt)}{c})}{\sigma_{\nu}} \right)^2 \right],
\end{equation}
where $\sigma_{\nu}$ gives the width of the spectral line. The line of sight velocity $\vL$ can be written in terms of the contribution from the tangential rotation velocity $\vR$ and the contribution from the random turbulent velocity $\vT$ as
$	\vL = \vR + \vT.$ 
For an exactly face on disk $\vR = 0$. We define the $j^{\rm th}$ moment of the specific intensity and the local average of it as
\begin{eqnarray}
\label{eq:Imom}
M_j (\vt) &=& \int \id \nu\, \nu^{j}\ I(\vt, \nu),  \\
\langle M_j (\vt)  \rangle &=& \int \id \vt' M_j(\vt') L(\vt - \vt'),
\end{eqnarray}
were $L(\vt)$ is the local averaging kernel. The convolution of $M_j(\vt)$ with  $L(\vt)$ brings out the
large scale features of $M_j(\vt)$. The quantities $\langle M_j (\vt)  \rangle$ gives  the $\vt$ dependence of the  large scale variation in the column density, i.e  $G(\vt)$ for $j=0$ and the line of sight component of the tangential rotational velocity $v^{\Omega}(\vt)$ with $j=1$. Note that the definition of $ M_j (\vt) $ used here does not correspond to the usual definition of the image moments.
\subsection{Visibility moments}
The directly observed quantity in a radio interferometric observation is  visibility $V(\vU, \nu)$. Visibilities are measured as  a function of the  baseline vector  $\vec{U}$, the projected  separation of antenna pairs in a plane perpendicular to the line of sight of observations, in units of the observing wavelength. Since the angular extent of the external galaxies we consider here are rather small, we can approximate the visibilities as Fourier transform of the sky brightness distribution, i.e,
\begin{equation}
V(\vU, \nu) = \int \id \vt e^{-i 2 \pi \vt . \vU}\, A(\vt) I(\vt, \nu),
\label{eq:defvis}
\end{equation}
 where $A(\vt)$ is the beam pattern of each antenna or the primary beam of the telescope. Here we assume that during the observation,  the antennas are tracking the galaxy, i.e, the origin of our coordinate system ($\theta_x, \theta_y, z$) is always at the centre of the field of view of observation.  In general, the antenna beam pattern is a function of the observing frequencies. However, for the observation of \HI emission from nearby external galaxies, the fractional bandwidth is small enough  and the variation of $A(\vt)$ with observing frequencies can be ignored. 
The $j^{\rm th}$ moment of the visibility can be written as
\begin{equation}
V_j (\vU) = \int \id \nu\, \nu^{j}\ V(\vU, \nu).
\end{equation}
Using the eqn~(\ref{eq:Imom}) we  define the followings
\begin{equation}
W_j (\vU) = \int \id \vt e^{-i 2 \pi \vt . \vU}\, A(\vt) \langle M_j (\vt)  \rangle.
\end{equation}
 The quantities $W_j (\vU) $ carries the information about the large scale variation of the specific intensity ($j=0$) and the line of sight component of the tangential rotational velocity ($j=1$) of the galaxy.

\subsection{Power spectrum of column density and velocity}
The power spectrum of the column density $P_{\rm HI}$ and the turbulent velocity  $P_{v}$  fluctuations are defined as
\begin{eqnarray}
\langle \delta \tilde{n} (\vU')^{*} \delta \tilde{n} (\vU)\rangle &=& \delta (\vU - \vU') P_{\rm HI} (\vU') \\
\langle \delta \tilde{v}^T (\vU')^{*} \delta \tilde{v}^T (\vU)\rangle &=& \delta (\vU - \vU') P_{v} (\vU').
\end{eqnarray}
Here $\delta \tilde{n}$ and $\delta \tilde{v}^{T}$ are the two dimensional Fourier transform  (as used in eqn~\ref{eq:defvis}) of the density and velocity fluctuations $\dn, \vT$ respectively.   The symbol ($^*$) denotes complex conjugate and $\delta(\vU)$ is the two dimensional Dirac delta function.

 The function $V_0(\vU)$ has the information of the large scale variation of the intensity across the sky, as well as the fluctuations in the column density, whereas the function $W_0(\vU)$ depends only on the large scale variation of the intensity across the sky. We define $X_0 (\vU) = V_0 (\vU) - W_0 (\vU)$, which gives the convolution of the column density fluctuation with $W_0(\vU)$, i.e,  
\begin{equation}
X_0 (\vU) =  W_0(\vU) \otimes \delta \tilde{n} (\vU).
\end{equation}
The symbol ($\otimes$) denotes convolution here.  We use $Q_0$ to define the following autocorrelation function of $X_0$:
\begin{eqnarray}
\label{eq:Q0def}
Q_0(\vU, \vU') &=& \langle X_0(\vU')^{*} X_0(\vU) \rangle \\ \nonumber
        &=& \int \id \vU_1 W_0(\vU - \vU_1 ) W_0(\vU '- \vU_1 ) P_{\rm HI}(\vU_1 ).
\end{eqnarray}
In the discussion of visibility based power spectrum estimator in \citet{2009MNRAS.398..887D} and in several earlier work \citep{Bharadwaj_2001}, it is demonstrated that such autocorrelation function decreases rapidly with increasing $\theta_0 \mid \vU - \vU' \mid$ for a galaxy with angular extent of $\theta_0$ in the sky. At baselines $\mid U \mid > 1/\theta_0$, we may write from eqn~(\ref{eq:Q0def})
\begin{equation}
\label{eq:PHIdef}
P_{\rm HI} (U) =  \lim_{\mid U \mid  \theta_0 \to\infty,  \theta_0 \mid \vU - \vU' \mid \to 0}  {} \frac{Q_0 (\vU, \vU') }{\int \id \vU' \mid W_0 (\vU') \mid ^2 }.
\end{equation}
Hence, this works as an estimator for the \HI column density fluctuations. Here we have also assumed that the fluctuations in the column density are statistically homogeneous and isotropic and hence are the only function of $U = \mid \vec{U}\mid$. 

To estimate the power spectrum of the turbulent velocity fluctuations, we proceed as follows. We define  $X_1 (\vU) = \frac{c}{\nu_0} \left [ V_1 (\vU) - W_1 (\vU) \right] $, where $c$ is the speed of light. $X_1 (\vU)$ can be written in terms of the column density and turbulent velocity fluctuations as
\begin{equation}
X_1 (\vU) = \frac{c}{\nu_0}\ W_1(\vU) \otimes \delta \tilde{n} (\vU) + W_0 (\vU) \otimes \delta \tilde{v}^{T} (\vU).
\end{equation}
The autocorrelation of $X_1$ gives
\begin{eqnarray}
        \label{eq:Q1def} 
 Q_1(\vU, \vU') &=& \langle X_1(\vU')^{*} X_1(\vU) \rangle \\ \nonumber
        &=&  \int \id \vU_1 W_0(\vU - \vU_1 ) W_0(\vU '- \vU_1 ) P_{v }(\vU_1 ) \\ \nonumber
        &+& \left ( \frac{c}{\nu_0}\right)^2 \int \id \vU_1 W_1(\vU - \vU_1 ) W_1(\vU '- \vU_1 ) P_{\rm HI}(\vU_1 ) 
\end{eqnarray}
 Here we have assumed that the standard deviation in the relative  fluctuations in the column density $\delta n$ is smaller than unity. We also assume that the standard deviation of the  line of sight component of the random turbulent velocity $\delta v^{T}$ is smaller than the line of sight component of the   tangential rotation velocity $v^{\Omega}$. We neglect the contribution in $Q_1(\vU, \vU')$ that arises from multiplication of $\delta v^{T}$ and $v^{\Omega}$. The power spectrum of \HI velocity fluctuations can be written as
\begin{equation}
\label{eq:Pvdef}
P_{v} (U) =  \lim_{\mid U \mid  \theta_0 \to \infty,  \theta_0 \mid \vU - \vU' \mid \to 0} {}  \frac{ \left [ Q_1 (\vU, \vU') - H(\vU, \vU') \right]}{ \int \id \vU' \mid W_0 (\vU') \mid ^2 }, 
\end{equation}
where $H(\vU, \vU')$ is the second term in the right hand side of eqn~(\ref{eq:Q1def}). 
Hence, this works as an estimator for the \HI velocity fluctuations.

\section{Observation and Analysis}
\subsection{Data and Calibration}
\label{DataAnalysis}
 The visibility moment estimators  (VME) discussed here assumes that the galaxy of interest has a nearly face on  thin  disk. A typical external spiral galaxy's disk is expected to have an arbitrary inclination angle with respect to the line of sight direction. The nearly face on disk ensures that the angular separation we observe can be directly related to length scales on the face of the disk. 
 Further, considering a tilted ring model for the galaxy's large scale dynamics, the inclination, as well as the position angle is required to be smoothly varying with  the radial separation from the galaxy's centre (galacto-centric distance).  The VME is originally discussed in \cite{2016MNRAS.456L.117D}. They used models of galaxies to show  that the VME for the power spectrum of the line of sight velocity  fluctuations has less uncertainty for inclination angle  between   $15^0$ and $40^0$.  Hence, the VME should be used only for the galaxies with disk inclination angle in this range.  In this work, we investigate turbulence in the disk galaxy NGC 5236 (M~83), a barred spiral galaxy. Distance to this galaxy is estimated in literature using the Tip of the Red Giant Branch (TRGB) \citep{Radburn_Smith_2011} and other luminosity based methods \citep{1992ApJ...395..366S,1994ApJ...430...53P}.  We adopt a distance of $4.77$ Mpc to the galaxy. The galaxy has a nearly face-on disk with an average  inclination angle of $24^{\circ}$  \citep{2008AJ....136.2563W,1993A&A...274..707T}. The tilted ring rotation curve model  from earlier \HI studies show that the inclination and position angles vary monotonically with  galacto-centric distance \citep{1981A&A...100...72H}. The galactic disk has a large angular extent of $30'\times 24'$ at a \HI column density of $10^{19}$ atom/cm$^2$ \citep{2013NewA...19...89D}, which makes it suitable for our study.
 
 NGC~5236 was observed with the uGMRT on 02, 03 June and 31 July 2018 for a total observation time of $14$ hours at the L band. We used a total of 4096 channels over a 12 MHz bandwidth centred at the redshifted frequency of the galaxy. The larger bandwidth was chosen  to accurately model the continuum emission. Each channel in the data corresponds to a velocity resolution of $0.6$ km sec$^{-1}$, adequate to study the turbulent velocity fluctuations in the galactic disk.
Primarily, we processed the observations of each day separately. We used the standard tasks in classic AIPS\footnote{AIPS: Astronomical Image Processing System, NRAO} to  edit the data  for  radio frequency interference and primary calibration. The calibrated data from each day is then combined using the AIPS task DBCON and exported to a FITS \footnote{FITS: Flexible Image Transport System} file.  The half power beam width of the uGMRT at L-band is $25'$. Given the large \HI extend of NGC~5236, it is required to use modern deconvolution algorithms to estimate the continuum model for this galaxy. 
We used the CASA\footnote{CASA: Common Astronomy Software Application, NRAO} for modelling the continuum and self-calibration. CASA task tclean provides better options for imaging in a large field of view observations. The continuum model is subtracted from the visibilities in each channel using the UVSUB algorithm in the CASA.

Angular extend of the NGC~5236 is rather large and it barely fits in the primary beam of the GMRT at L band. Further, at low baselines, the GMRT baseline coverage is rather poor resulting in less sensitivity at large scales. \citet{2008AJ....136.2563W} have used different array configuration of the VLA  to observe $34$ nearby galaxies including NGC~5236. The galaxy is observed with the VLA hybrid configurations BnA, CnB and DnC for about 12, 2 and 20 hours respectively.  Velocity resolution corresponding to each channel in  this observation is $2.6$ km sec$^{-1}$.  We use the  THINGS UV data  to improve on the sensitivity and baseline coverage at smaller baselines. 
The VLA CnB and DnC configurations complement to the GMRT observations, where the VLA observations provide better baseline coverage at the smaller baselines and the GMRT observation at larger baselines. We start with  the THINGS data with primary calibration \footnote{We are indebted to Prof. Fabian Walter 
for providing us with the calibrated uvdata.} and use the same procedure as discussed above to produce the continuum subtracted UV data for further analysis.  The GMRT and VLA observations are separately processed with our power spectrum estimators. 

\subsection{Implementation of the Visibility Moment Estimator}
\label{implementation}
\citet{2019RAA....19...60D} show that the scale-dependent quantities like the power spectrum of the specific intensity fluctuations can be estimated unbiasedly directly in the visibility domain. They also show that the large scale structure of the galaxy, like the $\langle  M_j (\vt) \rangle $ can be estimated from the reconstructed image. We estimate the first and second moment of visibility $V_0(\vU)$ and $V_1(\vU)$ from the calibrated and continuum subtracted visibilities. Since in a radio interferometric observation, the visibilities are sampled at random positions in the baseline plane, the visibility moments are estimated only at the sampled baseline positions. Note that the baselines for a given antenna pair at a given time are different for different frequency channels. However, owing to the small fractional bandwidth,  the difference in our observation is rather small and can be ignored. We generate position-position-frequency data cube from the continuum subtracted visibilities and use it to estimate  $M_j(\vt)$.  A Gaussian kernel $L(\vt) = \frac{1}{2 \pi \theta_0^2} \exp (-\mid \vt \mid^2/2 \theta_0^2)$ with $\theta_0 = 5'$ is used to estimate the $\langle  M_0 (\vt) \rangle $ and $\langle  M_1 (\vt) \rangle $. For a Gaussian approximation to $M_j (\vt)$, this value of $\theta_0$ gives a column density of $10^{19}$ atoms cm$^{-2}$ at  $4 \times \theta_0$.  We calculate the discrete Fourier transform of the $M_j$ multiplied with the primary beam of the corresponding telescopes at the sampled baseline positions to estimate $W_j(\vU)$. The corresponding $X_j(\vU)$ are estimated at the sampled baseline positions by direct subtraction of the $W_j(\vU)$s from the visibility moments.

\citet{2013MNRAS.436L..49D} and \citet{2013NewA...19...89D} used a visibility based power spectrum estimator to calculate the power spectrum of \HI intensity fluctuations from nearby galaxies. The power spectrum estimator uses visibility correlation at nearby baselines to reduce possible bias that may arise from the measurement noise. In this work, we use a similar visibility correlation estimator for quantity $Q_0$. To estimate the power spectrum at a baseline $\vU$, we first  correlate every  pair of $X$ for all baselines within a circular region of radius $D$ in the baseline plane centred at $\vU$.  The value of  $D$ need to be  smaller than the inverse of the  angular extent of the galaxy (see eqn~\ref{eq:PHIdef} and eqn~\ref{eq:Pvdef}). 
We then average the correlations in each circular regions (as stated above) in annular bins in the baseline plane.  To calculate the column density power spectrum $P_{\rm HI}(U)$, we proceed to estimate the denominator in eqn~(\ref{eq:PHIdef})
 in the following way. We note that using the properties of Fourier transform, $\int \id \vU' \mid W_0 (\vU') \mid ^2 = \int \id \vt' \mid \langle M_0 (\vt') \rangle A(\vt) \mid ^2$, and use the later for estimation of the denominator. 
The real part of the ratio of the annular averaged quantity discussed above to the calculated denominator gives estimates of the $P_{\rm HI}$ at baselines larger than $1/\theta_0$.  The visibility correlation at nearby baselines reduces the amplitude of the correlation with respect to the correlation at the same baselines \citep{2013NewA...19...89D}.  The effect of this bias is larger if a relatively larger value is chosen for $D$. The imaginary part of the same quantity measures a bias in the above estimator that may arise  if $D$ is chosen to be large. Hence, the ratio of the real to the imaginary part at each baseline gives a diagnostics on the choice of $D$. We check the ratio of the real to the imaginary part at each baselines and only consider points where the ratio is more than $3$.  \citet{2011arXiv1102.4419D} discuss the errors in such estimators. The errors we quote here are calculated using the same method assuming that the error in the estimate of the denominator in eqn~(\ref{eq:PHIdef}) is negligible. 
 
 To estimate the velocity power spectrum $P_v$ we proceed the following way. At a baseline $\vU$, we choose all other baselines $\vU'$ within a circular region of radius $D$ as before and  calculate the function $H(\vU, \vU')$ using our estimate of the $P_{HI}$ and $W_1$. 
 For each of these pair of baselines, we also calculate $X_1(\vU')^{*} X_1(\vU)$ and subtract $H(\vU, \vU')$ from it. The real part of the azimuthal average of this quantity normalised by the $\int \id \vU' \mid W_0 (\vU') \mid ^2$ gives estimates of  the velocity power spectrum at baselines larger than $1/\theta_0$.  Errors in this estimators are calculated in the similar way as for the column density power spectrum assuming the errors in $W_1$ is negligible. Further, we also observed that the errors in the column density power spectra are rather small and do not contribute much to the errors in the velocity power spectrum.
 
\section{Result and Discussion}
 We process the GMRT and the VLA data discussed in section~\ref{DataAnalysis} separately for the power spectrum analysis. 
The \HI emission for the galaxy NGC~5236 spans over a velocity range of about $~300$ km sec$^{-1}$  \citep{2008AJ....136.2563W}.  The angular extent of  the galaxy NGC~5236 is $24' \times 30'$. A value of $\theta_0 \sim 25 '$ corresponds to a baseline of $\sim 0.15$ k$\lambda$ \footnote{k$\lambda$ means 1000 wavelengths.}. Conservatively, for the baseline range $> 0.3 $ k$\lambda$ we may use eqn~(\ref{eq:PHIdef}) and eqn~(\ref{eq:Pvdef}) to estimate the column density and velocity power spectra respectively.  We use $D=0.075\  {\rm k}\lambda$  for our analysis.   For the GMRT observations we choose the central $200$ channels that have a  comparatively higher signal to noise. These correspond to a velocity width of $120$ km sec$^{-1}$. Note that restricting out analysis only to the central channels does not affect the measurement of the power spectra, it only limits the measurements of the column density and velocity at a limited part of the baseline plane. This was used as an advantage by \citet{2010MNRAS.405L.102D} to estimate the power spectra of different parts of the galaxy NGC~4254.  We follow the procedure highlighted in section~\ref{implementation} to estimate the column density power spectra and the corresponding errors for the GMRT observations. 
\begin{figure}
\begin{center}
\includegraphics[scale=0.45]{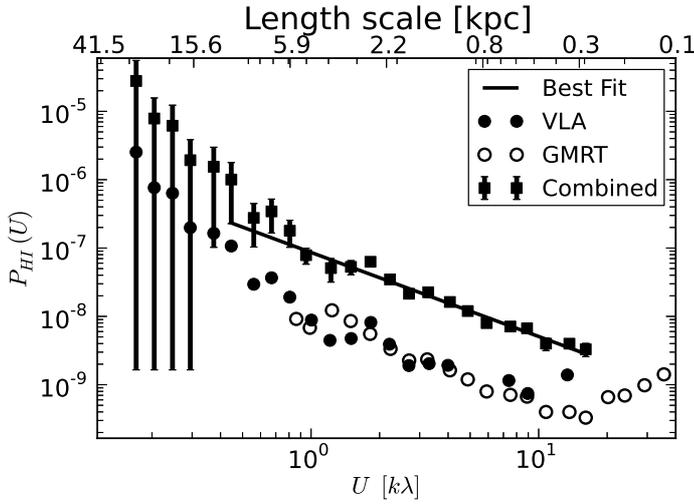}
\caption{Column density power spectrum of NGC~5236 as a function of baseline. The corresponding length scales are also shown in the top margin. The black solid circles correspond to the measurement of the column density from the VLA observations alone. The black open circles correspond to measurement using only the uGMRT observations. These two sets of points are scaled down by a factor of ten for display purpose. The black solid squares with error bars gives the power spectra measurements combining the VLA and the uGMRT observations. A best fit power law to the combined measurement is shown using black solid lines.}
\label{fig:PHI}
\end{center}
\end{figure}
The empty circles in Figure~\ref{fig:PHI} shows the measurements of the azimuthally averaged values of $P_{\rm HI}$ as a function of baselines. We show here points where the real part of the right hand side of eqn~\ref{eq:PHIdef} is more than its imaginary parts. 

We use the central $90$ channels of the VLA data with a higher signal to noise for the power spectrum analysis. The estimated values of the column density power spectrum is shown with black filled circles in  Figure~\ref{fig:PHI}. We show here only the points with  the real part of the right-hand side of eqn~\ref{eq:PHIdef} greater than its imaginary part. The power spectrum values for both the GMRT and VLA observations are scaled down by a factor of $10$ from its original values for presentation purpose. 

Clearly, at baselines smaller than $\sim 0.8$ k$\lambda$ the GMRT observation do not provide any measurement of the power spectrum.   The  VLA observation gives better estimates of the power spectrum at lower baselines, however, at larger baselines, the estimates are dominated by measurement noise. We fit a power law of form $A\ U^{\alpha}$ to the data and estimate the best fit parameters for both the GMRT and the VLA observations separately. The range of fit in baselines, the corresponding length scales at the galaxy, the value of the reduced chi-square and the best fit values with $1-\sigma$ errors for the parameters $A$ and $\alpha$ are given in Table~1.  
\begingroup
\begin{table}
	
\setlength{\tabcolsep}{1pt} 
\renewcommand{\arraystretch}{.8}
\begin{tabular}{|c|c|c|c|c|}
\hline
  &
$P_{HI}(\vec{U})$ & $P_{HI}(\vec{U})$ &
$P_{HI}(\vec{U})$ &
$P_{v}(\vec{U})$\\ \hline \hline
Telescope & VLA & GMRT & Combined & GMRT \\  \hline 
$A (\times 10^{4}) $                  & $4.0 \pm 0.8$ & $5.0 \pm 0.6$ & $4.0 \pm 0.4 $ & $88 \pm  9 ^{*}$ \\ \hline
$\alpha$             & $-1.2 \pm 0.3 $             & $-1.24 \pm 0.07$              & $-1.23 \pm 0.06$               & $-1.91 \pm 0.08$ \\ \hline
reduced $\chi^2$ & 1.7& 1.1& 1.1& 1.1\\ \hline
$U_{min} (K\lambda) $& 0.6                          & 0.9                          & 0.4                          & 0.9             \\ \hline
$U_{max} (K\lambda) $& 3                          & 16                         & 16                         & 16            \\ \hline
$R_{min} (Kpc)$      & 1.4                          & 0.3                          & 0.3                           & 0.3             \\ \hline
$R_{max} (Kpc)  $    & 9                          & 5                         & 11                          & 6            \\ \hline
\label{tab:tab1}
\end{tabular}
\caption{Result of power law fit ($P = A U^{\alpha}$) to the column density ($P_{\rm HI}$) and velocity ($ P_v$) power spectra. The power law amplitude ($A$) at one steradian, best fit slope $\alpha$, $1-\sigma$ errors associated with the fit, the reduced $\chi^2$ values and the range of fit in baselines and corresponding length scale ranges are shown. The density power spectra fit are shown for the THINGS VLA data, the GMRT data and the combined power spectra. The velocity power spectra results are only from the GMRT data. The amplitudes for the density spectra are scaled by a factor of $10^4$ and have a unit of steradian. $^*$The amplitude of the velocity spectra are not scaled and is given in (km sec$^{-1}$) $^2$ steradian. }
\end{table}
\endgroup

 It is to be noted that since we choose the channels with a higher signal to noise only to estimate the power spectra, the effective angular extent of the galaxy reduces. As discussed earlier, this results in a more stringent choice in the value of $D$ and may  bias the visibility moment estimator near to the inverse of the effective angular scale. To avoid this bias  we only use the estimates with the real part three times or more than the imaginary part. 
 Black squares with error bars show the combined power spectrum from the VLA and the GMRT estimates (there is no scaling of these values). We use the following procedure to combine the power spectra from the GMRT and the VLA observations. We first choose the union of the baseline ranges over which the individual power spectra are estimated and divide it into $N_{bin}$ equal logarithmic bins. We average the estimate of the power spectra in each bin after weighting them by the corresponding errors. The error bars in each bin are calculated by taking the root mean square of the  individual error estimates in the given bin.
 It can also be seen clearly in Figure~\ref{fig:PHI} that at smaller baselines the power spectra ( solid squares) deviates from the power law behaviour and excess power is measured. This is the region where $\mid \vU \mid < 1/\theta_0$ and the convolution effect in eqn~(13) is important. We find that  the combined power spectra assume a power law between the baseline range of $0.4 - 16$ k$\lambda$. The best fit value of $A$ and $\alpha$ to the combined power spectra are $(4.0 \pm 0.4)\times 10^{-4}$ steradian and $-1.23\pm0.06$ with the goodness of fit given by the reduced chi-square of $1.1$ (see Table~1).  The amplitudes mentioned here are at the baseline of one wavelength. This is the first determination of the column density power spectra over almost two decades of  length scale  between $300$ pc to $11$ kpc. 

\begin{figure}
\begin{center}   
\includegraphics[scale=0.45]{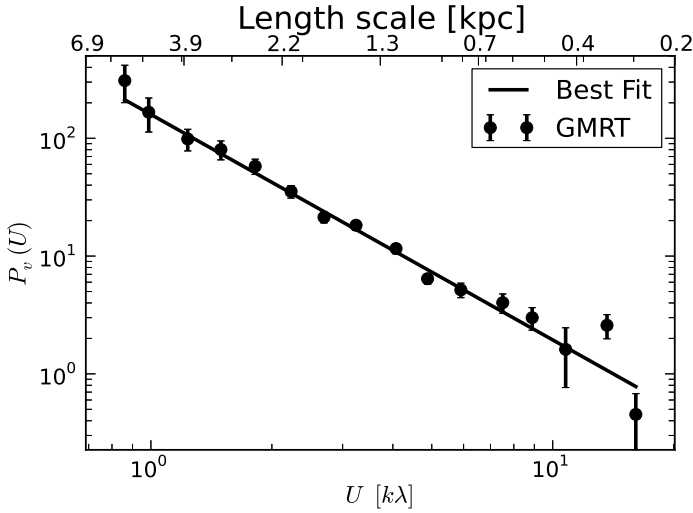}
\caption{The turbulent velocity power spectrum of NGC~5236 as a function of baseline. The corresponding length scales are also shown in the top margin. The black solid circles with error bars correspond to the measurement from the  uGMRT observations. A best fit power law to the combined measurement is shown using black solid lines.}
\label{fig:Pv}
\end{center}
\end{figure}

 We use a similar procedure to estimate the power spectrum of the line of sight velocity fluctuations using the GMRT observation. The VLA observation has relatively low velocity resolution of $2.6$ km sec$^{-1}$ compared to what we would have required to estimate the velocity power spectrum. We do not use the VLA data here. The power spectrum estimates along with the error bars are shown with black  circles in Figure~\ref{fig:Pv}. We find that the power spectra can be well fitted by a power law in the baseline range $0.9$ k$\lambda$ to $16$ k$\lambda$, with a goodness of fit value of $1.1$ given by the reduced chi square. The best fit power law gives   the amplitude of $ P_v$ as $88 \pm 9$ (km sec$^{-1})^2$ steradian. The slope of the best fit power law is found to be $-1.91\pm0.08$. This is the first determination of a power law power spectrum of line of sight velocity fluctuation for an external spiral galaxy.

\section{Conclusion}
In this work, we measure \HI column density and line of sight velocity power spectra of the galaxy NGC~5236 using visibility moment estimators with  the GMRT and the VLA observations. The power spectra are estimated over almost two decades of length scales ranging from $\sim 300$ pc to $11$ kpc on the galaxy's disk.  These are the first measurements of the density power spectra on such large length scale ranges  and the first measurement of the velocity power spectra of any external spiral galaxy.

Using the combined data from the GMRT and VLA observations described earlier, we find that the column density power spectrum of the galaxy NGC~5236 to follow a power law over a length scales ranging $300$ pc to $11$ kpc with a slope of $-1.23\pm0.06$. \citet{2013NewA...19...89D} reports that for $10$ of the $18$ external spiral galaxies from the THINGS survey, the slope of the specific intensity fluctuation power spectra lies within $-1.5$ to $-1.9$ over a  length scale range of $300$ pc to $16$ kpc across all the galaxy sample.  For the galaxy NGC~5236, the   power spectrum  follows a power law with a slope $-1.9\pm0.2$ between the baseline range of $ 0.8 - 7.5$ k$\lambda$ using the VLA observations from the THINGS survey. The mean value of the slope of the power spectrum we find here with the  VLA data is different than their result, however, the slopes are consistent within three sigma uncertainties (see Table~1). It is also to note that our estimates of the slope and the amplitudes of the power spectra individually from the GMRT and the VLA observations  are mutually consistent. 

\begingroup
\setlength{\tabcolsep}{3pt} 
\renewcommand{\arraystretch}{.8} 
\begin{table}
\begin{tabular}{|lccc|}
\hline
Type of \\ Turbulence   & $P_{\rm HI}(\vec{U})$  & $P_{v}(\vec{U})$  & References  \\ \hline \hline
Kolmogorov  		& -                   & $-5/3$            & \citet{1941DoSSR..30..301K}\\ 
Burger's    		& -                   & $-2$              & \citet{BURGERS1948171}\\
	Weizsacker              & $6\alpha - 1$       & $-5/3 - 2\alpha$  & \citet{1996ApJ...458..739F} \\
Solenoidal  		& $-0.78\pm0.06$      & $-1.86\pm0.05$    & \citet{2009ApJ...692..364F}\\ 
Compressive 		& $-1.44\pm0.23$      & $-1.94\pm0.05$    & \citet{2009ApJ...692..364F}\\ 
NGC~5236		& $-1.23\pm0.06$      & $-1.91\pm0.08$    &  This work\\ \hline
\end{tabular}
\caption{Power spectral slope for density and velocity for different theoretical models of turbulence and simulations. All the spectral slopes are given for what is expected for turbulence in the thin disk as the case for this observational result.}
\label{tab:comp}
\end{table}
\endgroup

Our measurement that  the column density and the velocity power spectra assume  power laws over large range of length scales  strongly suggests about the presence of large scale turbulence cascade in the galaxy's disk.  \citet{10.1111/j.1745-3933.2009.00684.x} show that for the galaxy NGC~1058 the specific intensity power spectra steepens at a length scale corresponding to 1.5 kpc indicating a transition from the two-dimensional structure at the large scale to the three-dimensional structures at scales smaller than the scale-height of the galaxy.   \citet{2010ApJ...718L...1B} and \citet{, 2012A&A...539A..67C} observe similar trends in the power spectra of the dust emission in the LMC and the nearby galaxy M~33 respectively. The fact that we do not observe any such steepening down to a  scale of $300$ pc, suggests that the scale height of NGC~5236 is less than this scale. Hence, the power law slope we see here can be interpreted as arising from two-dimensional turbulence on the galaxy's disk.

Table~\ref{tab:comp} compares our result about the slope of the density and velocity power spectra of \HI in the galaxy NGC~5236 with different theoretical models and simulations of turbulence.  The \citet{1941DoSSR..30..301K} universal equilibrium theory for incompressible fluid turbulence predicts a velocity power spectrum with slope of $-5/3$ for the galaxy's disk geometry we are discussing here.  The mathematical model by \citet{BURGERS1948171} of compressive turbulence flows gives a slope of $\sim -2$ for the velocity power spectrum. \citet{2009ApJ...692..364F} perform numerical simulations of compressible fluid turbulence and investigate the role of solenoidal and compressive forcing. They find for a pure solenoidal forcing the density and the velocity spectra assume slopes of $ -0.78\pm0.06$ and $-1.86\pm0.05$ respectively. In the case of compressive forcing, the density power law slope is lower $-1.44\pm0.23$, whereas the velocity spectrum slope is almost similar as before $-1.94\pm0.05$. Our measurement of the density and velocity power spectral slopes of $\sim-1.23$ and $\sim-1.91$ suggests that most of the forcing behind the large scale ISM turbulence we probe here is  compressive in  nature. Using the phenomenological model of compressive turbulence by \citet{1951ApJ...114..165V}, \citet{1996ApJ...458..739F} commented on the relation between the velocity and column density scaling. Using this suggestion,   the velocity and column density fluctuations in compressible turbulence is related by scaling relations as given in Table~\ref{tab:comp}. Following this prescription, for a column density scaling of $-1.23$, we get a velocity power spectrum scaling of $-1.92$, consistent with the measured value here.

Our measurement gives a column density fluctuation power law amplitude of $4\times 10^{-4}$ at an angular scale of one steradian. For a power law slope of $-1.23$,  the expected amplitude of the autocorrelation function of the column density distribution is $0.77$. This corresponds to a standard deviation in column density fluctuation of $0.011$ at a length scale of $11$ kpc. 

\begin{figure}
\includegraphics[scale=.5]{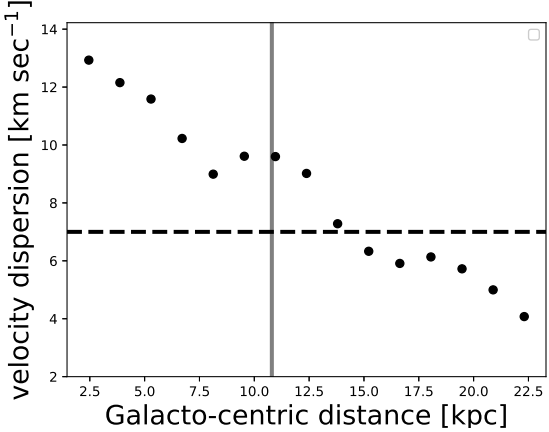}
	\caption{Median velocity dispersion is plotted with galacto-centric radius (circles). The vertical line corresponds to the $r_{25}$ of the galaxy. The horizontal line corresponds to the turbulence velocity dispersion at $6$ kpc scale.}
\label{fig:vdisp}

\end{figure}

 \citet{2010MNRAS.409.1088B} have investigated the role of self-gravity, turbulent cascade and stellar feedback in the properties of the ISM in disk galaxies using numerical simulations. They find the turbulence at the large scale is insensitive to the stellar feedback and mostly driven by gravitational instabilities. They conclude that the origin of the scale-invariant motions at the large scales is a possible result of a forward cascade from turbulent energy input at large scale from the spiral waves and other long range disturbances combined with an inverse cascade from the small scales. \citet{2003ApJ...593..333E} investigate the optical structure of M~33 at three different passbands. They conclude that the distribution of stars follows a fractal pattern generated by the background distribution of the turbulent ISM. 
\citet{2016AJ....152..134M} use the sample of galaxies in the THINGS survey to investigate the relationship between turbulence and star formation. They conclude that the large scale turbulence is more likely to be driven by disk gravitational instabilities, density waves or bar streaming motions and is less likely to be results of stellar activities. \citet{2012AJ....144...96I} find that at the outer part of the galaxy's disk, the \HI velocity dispersion has a broad component  with velocity dispersion in excess of $\sim 5$ km sec$^{-1}$, suggesting that the star formation is not the major player to drive the \HI velocity dispersion favouring turbulence driving at the exteriors of the galaxy's disk. Similar conclusion is received by the study by \citet{2019ApJ...887..111S} and \citet{2017ApJ...845...53N} on the Large Magellanic Cloud, where they conclude that there is significant turbulence driving exists at the large scales from sources other than stellar feedback. {\citet{2016MNRAS.458.1671K} also show that it is more likely that the gravity is the source of the large scale ISM turbulence and wins over the stellar feedback.  \citet{2003ApJ...590..271E} mention about a turbulent origin of the spiral structures in the galaxies.  They compare the size of the largest structures with the critical Jeans length at that scale. They conclude   that the self gravity driven shear instabilities in large-scale   generates the  flocculent spiral arms. The energy of these instabilities drives turbulence at these scales  which follow a forward cascades to form smaller scale structures.

 We find that the turbulent velocity fluctuations in the \HI have a power of  $88$ (km sec$^{-1}$)$^2$ at unit steradian. The corresponding  power spectrum slope of $-1.91$ gives a power law slope of $0.09$ in the autocorrelation function of the turbulent velocity fluctuations. Assuming the power law behaviour, at the scale of $6$ kpc, the amplitude of the autocorrelation function of turbulent velocity fluctuation is $\sim 48.3$ 
(km sec$^{-1}$)$^2$. This gives a turbulent velocity dispersion of  $\sim 7$ km sec$^{-1}$ at $6$ kpc scales. Considering a number density of $1$ atom cm$^{-3}$, a velocity dispersion of $7$  km sec$^{-1}$ corresponds to a Jeans  scale of $\sim 4.5$ kpc.  In this work, the largest length scale to which we observe the power spectrum to  follow power law is $11$ kpc for the column density power spectra and $6$ kpc for the velocity power spectra. Hence, the estimated Jeans scale is quite comparable to the largest scale to which we probe the power law velocity spectra. We believe this as an evidence that the  gravitational instability drives the observed turbulence at $\sim 6$ kpc scales and maintains the large scale power spectra. \citet{2014MNRAS.442.1230R} discussed the importance of gravitational instabilities in generating coherent  structures and it's influence on star formation in  high redshift galaxies. Using a Toomre-like stability criteria \citep{2010MNRAS.407.1223R} of the gas disks with  the CO and \HI velocity dispersion of a few nearby galaxies \citet{2017MNRAS.469..286R} show that the characteristics scale  for disk instability is about $6$ kpc, quite similar to what we observe here.

 We use the natural weighted moment 2 map of NGC~5236  from the THINGS archive  \footnote{http://www.mpia.de/THINGS} to estimate the radial variation of the \HI velocity dispersion. Median velocity dispersion for NGC~5236 in elliptical annular regions are plotted against the galacto-centric distance in Figure~\ref{fig:vdisp}.  The ellipse position angle is chosen as the average position angle of the galaxy's disk  while the ellipticity is derived from the average inclination angle of the disk $225^{\circ}$ and $24^{\circ}$ respectively \citep{2008AJ....136.2563W,1993A&A...274..707T}.  The vertical line corresponds to the r$_{25}$ of the galaxy (\citet{2008AJ....136.2563W} table 1). The horizontal dashed line corresponds to the velocity dispersion at 6 kpc scale due to turbulence. Clearly turbulence plays a significant role in the total velocity dispersion and is very significant to drive the dispersion beyond the stellar disk of the galaxy. The velocity dispersion drops sharply beyond 12 kpc. Interestingly, this is twice the scale to which we probe the turbulence.

The energy input rate per unit area by the ISM turbulence can be estimated using 
$\epsilon = \frac{1}{2} \times (N_{{\rm HI}_0} \sigma_{\rm HI} m_{\rm HI} )\times (v^{T})^2 \times  v^{T}/L$,
where $N_{{\rm HI}_0}$ is the average column density over the disk, $m_{\rm HI} $ is the mass of atomic hydrogen,   $\sigma_{\rm HI}$ is the relative fluctuation of \HI column density and $v^{T}$ is the turbulent velocity fluctuation. The quantity $N_{{\rm HI}_0} \sigma_{\rm HI} m_{\rm HI}$ gives the mass of \HI in turbulence per unit area, $L/v^T$ gives the time scale of energy input. We find from the moment 0 map of the galaxy NGC~5236,  the value of $N_{{\rm HI}_0} = 4 \times 10^{20}$ atoms cm$^{-2}$. Considering the energy input scale is at $6$ kpc, we find the turbulence energy input is $\sim 7 \times 10^{-11}$ ergs cm$^{-2}$ sec$^{-1}$.
Considering an average energy released in a supernova of $10^{46}$ ergs as kinetic energy, a supernovae rate of one in 100 years, the average energy input rate by supernovae  per unit area in this galaxy is about $3.5\times 10^{-10}$ ergs cm$^{-2}$ sec$^{-1}$, quite comparable to the energy input in turbulence.  This may explain, why in spite of different energy input mechanisms at different scales, the power spectra of the ISM turbulence follow a two-component power law over a large range of length scales.

In this work, we conclude the existence of energy input into the ISM turbulence at scales as large as $6$ kpc. The forcing is more compressive in nature. The Jeans scale estimated from the density and velocity fluctuations at these scale suggests that the energy include may be from the  gravitational instability or self-gravity in the disk. The observed turbulence cascade lies in the galaxy's disk and it continues till a length scale of at least $300$ pc.    \citet{2009ApJ...692..364F} show that incorporation of self-gravity and compressive forcing increases the density fluctuations in the ISM, which results in one order of magnitude higher star formation rates.  Out result suggests that the large scale turbulence driving can be a major player in  the global star formation rates of a galaxy. A possible way to investigate this further is to check  how the star formation rates are correlated with the velocity and density statistics of a sample of spiral galaxies. We plan to look into this in future work.

\section*{Acknowledgement}
MN acknowledge Department of Science and Technology - Innovation in Science Pursuit for Inspired Research (DST-INSPIRE) fellowship for funding this work. PD acknowledge DST-INSPIRE faculty fellowship support for this work.  
MN and PD thank the staff of the GMRT that made these observations possible. GMRT is run by the National Centre for Radio Astrophysics of the Tata Institute of Fundamental Research.
Authors thank the anonymous referee for suggestions that have improved the presentation of the paper significantly.	
\bibliographystyle{mnras}
\bibliography{main}

\end{document}